\documentclass[aps,prd,showpacs,superscriptaddress,floatfix,cite,twocolumn]{revtex4-2}
\usepackage{amsmath,amssymb,bm,color,fancyhdr,hyperref,lastpage,units,soul}
\usepackage{graphicx}

\usepackage[latin1,utf8]{inputenc}
\usepackage[english]{babel}
\usepackage{makecell}

\newcommand{\muB}{\mu_B}
\newcommand{\mI}{\mu_I}

\newcommand{\tr}{\text{tr}}

\newcommand{\D}{\bar{D}}
\newcommand{\dP}{\Delta P}
\newcommand{\mT}{\left(\frac{\mI}{T}\right)}
\newcommand{\Ns}{N_\sigma}
\newcommand{\Nt}{N_\tau}
\newcommand{\Nrv}{{N_\text{rv}}}
\renewcommand{\O}{\mathcal{O}}

\newcommand{\N}{\mathcal{N}}
\newcommand{\C}{\mathcal{C}}

\renewcommand{\N}{\mathcal{N}}
\newcommand{\trO}{\tr\,\O}
\newcommand{\second}{2^\text{nd}}
\newcommand{\fourth}{4^\text{th}}
\newcommand{\sixth}{6^\text{th}}

\def\lsim{\raise0.3ex\hbox{$<$\kern-0.75em\raise-1.1ex\hbox{$\sim$}}}
\def\gsim{\raise0.3ex\hbox{$>$\kern-0.75em\raise-1.1ex\hbox{$\sim$}}}

\begin{document}
\title{New Way to Resum the Lattice QCD Taylor Series Equation of State at Finite Chemical Potential}

\author{Sabarnya Mitra}
\affiliation{Centre for High Energy Physics, Indian Institute of Science, Bangalore 560012, India}

\author{Prasad Hegde}
\affiliation{Centre for High Energy Physics, Indian Institute of Science, Bangalore 560012, India}
\email[]{prasadhegde@iisc.ac.in}

\author{Christian Schmidt}
\affiliation{Fakult\"at f\"ur Physik, Universit\"at Bielefeld, Bielefeld D-33615, Germany}

\begin{abstract}
    Taylor expansion of the thermodynamic potential in powers of the (baryo)chemical potential $\mu_B$ is a well-known method to bypass the Sign Problem of Lattice QCD. Due to the difficulty in calculating the higher order Taylor coefficients, various alternative expansion schemes as well as resummation techniques have been suggested to extend the Taylor series to larger values of $\muB$. Recently, a way to resum the contribution of the first $N$ charge density correlation functions $D_1,\dots,D_N$ to the Taylor series to all orders in $\muB$ was proposed in Phys. Rev. Lett. 128, 2, 022001 (2022). The resummation takes the form of an exponential factor. Since the correlation functions are calculated stochastically, the exponential factor contains a bias which can be significant for large $N$ and $\muB$. In this paper, we present a new method to calculate the QCD equation of state based on the well-known cumulant expansion from statistics. By truncating the expansion at a maximum order $M$, we end up with only finite products of the correlation functions which can be evaluated in an unbiased manner. Although our formalism is also applicable for $\muB\ne0$, here we present it for the simpler case of a finite isospin chemical potential $\mI$ for which there is no Sign Problem. We present and compare results for the pressure and the isospin density obtained using Taylor expansion, exponential resummation and cumulant expansion, and provide evidence that the absence of bias in the latter actually improves the convergence.
\end{abstract}

\date{\today}
\maketitle

\section{Introduction}
\label{sec:introduction}
The Equation of State (EoS) of strongly-interacting matter is an important input in the hydrodynamical modeling of heavy-ion collisions~\cite{Bernhard:2016tnd,Parotto:2018pwx,Monnai:2019hkn,Everett:2020xug}. Unfortunately Lattice
QCD, which is the preferred method of calculating observables in the non-perturbative regime of QCD, breaks down when the baryon chemical potential $\mu_B$ is 
non-zero. This is the well-known Sign Problem of Lattice QCD~\cite{Nagata:2021bru}; despite recent progress~\cite{Aarts:2009yj,
Cristoforetti:2012su,Aarts:2013uxa,Sexty:2013ica,Fukuma:2019uot}, the current state-of-the-art results for the QCD EoS have been obtained by using either
either analytical continuation from imaginary to real $\mu_B$~\cite{Borsanyi:2018grb,Ratti:2018ksb}, or by expanding the EoS in a Taylor series in the chemical potential $\muB$ and calculating the first $N$ coefficients~\cite{Bazavov:2017dus,Bollweg:2021vqf}. 
In the latter case, a knowledge of the first several coefficients is necessary, not only to obtain the EoS for a fairly wide range of chemical
potentials but also to determine the radius of convergence of the Taylor series beyond which the Taylor expansion must break down~\cite{Gavai:2008zr, Dimopoulos:2021vrk, Bollweg:2022rps, Giordano:2019slo}. Unfortunately, the calculation of the higher order Taylor coefficients is computationally very challenging and it is natural to ask whether something can be learned about them from a knowledge of the first few Taylor coefficients. It turns out that this is indeed possible because 
the first $N$ derivatives $D_1,\dots,D_N$ of $\ln\det M(\muB)$, where $M(\muB)$ is the fermion matrix, also contribute to the higher order Taylor coefficients through products such as $D_N^2$, $D_ND_1$, etc. In fact, the contribution of the $n$th derivative $D_n$ to all higher orders can be shown to take the form of an exponential $\exp(D_n\muB^n/n!)$~\cite{Mondal:2021jxk}. Thus, if $D_n$ is known exactly, then its contribution to the Taylor series can be resummed to all orders through exponentiation. Exponential resummation, as we will refer to it from here on, can be shown to have several advantages compared to the original Taylor series: First, the resummed EoS converges faster than the Taylor series. Moreover, since the odd derivatives $D_1,D_3,\dots$ are purely imaginary, the resummed expression directly gives us a phase factor whose expectation value approaches zero as $\muB$ is increased, leading to a breakdown of the calculation. This breakdown is physical and related to the presence of poles or branch cut singularities of the QCD partition function in the complex $\muB$ plane. The resummed expression for the partition function also makes it possible to calculate these singularities directly. Some of these advantages have been recently demonstrated through analytical calculations in a low-energy model of QCD~\cite{Mukherjee:2021tyg}.

Despite its advantages, a technical drawback of exponential resummation is that the derivatives $D_1,\dots,D_N$ are not known exactly in an actual Lattice calculation. As is easily seen from the identity $\ln\det M = \tr\ln M$, the $D_n$ can be expressed in terms of traces of various operators, all of which involve the inverse of the fermion matrix $M$. Since $M$ is typically of size $10^8$ or greater, calculating its exact inverse is prohibitively expensive. Instead the various traces, and hence the derivatives $D_n$, are estimated stochastically using $\O(10^2$-$10^3)$ random volume sources per gauge configuration. Now, the products of such stochastically estimated quantities e.g. $D_N^2$, need to be evaluated in an unbiased manner \emph{i.e.} estimates coming from the same random vector must not be multiplied together. If $D_N^{(i)}$, $i=1,2,\dots,\Nrv$ are the $\Nrv$ stochastic estimates of the trace $D_N$, then the Unbiased Estimate (UE) of $D_N^2$ is given by
\begin{equation}
    \text{UE}\left[D_N^2\right] = \frac{2}{\Nrv(\Nrv-1)} \sum_{i=1}^{\Nrv}\sum_{j=i+1}^{\Nrv} D_N^{(i)}D_N^{(j)}.
\label{eq:unbiased_square}
\end{equation}
By contrast, the na\"ive Biased Estimate (BE) is given by
\begin{equation}
    \text{BE}\left[D_N^2\right] = \left[\frac{1}{\Nrv}\sum_{i=1}^{\Nrv}D_N^{(i)}\right]^2.
\label{eq:biased_square}
\end{equation}
Eqs.~\eqref{eq:unbiased_square} and \eqref{eq:biased_square} can both be readily generalized to any finite power or to the product of a finite number of traces.
In the Appendix, we present formulas for evaluating the unbiased estimate of such finite products in an efficient manner.
However we do not know of any corresponding formula to calculate the unbiased estimate of an infinite series such as an exponential. 

In this paper, we will present a new way of calculating the QCD EoS based on the well-known cumulant expansion from statistics. The cumulant expansion method is intermediate between a strict Taylor series expansion and exponential resummation in the sense that the contribution of $D_1,\dots,D_N$ are resummed 
only up to a maximum order $M$. However, since the order is finite it is possible to evaluate the terms of the expansion in an unbiased manner. 
The cumulant expansion agrees exactly with the Taylor series expansion to $\O(\muB^N)$ provided that 
$M \ge N$~\footnote{$N$ is assumed to be even from here on.}. However, it also contains additional
contributions at $\O(\muB^{N+2},\dots,\muB^{MN})$ which are exactly the contributions of $D_1,\dots,D_N$
to the higher order Taylor coefficients $\chi_{N+2}^B,\dots,\chi^B_{MN}$. 

Although the cumulant expansion method also works for $\muB\ne0$, in this paper we will present the formalism for the simpler case of finite isospin chemical 
potential $\mI$ instead. For $\mI\ne0$, the fermion determinant is real and one has no Sign Problem. Thus one only works with real quantities which in turn
simplifies the presentation. Moreover, the absence of the Sign Problem allows us to calculate observables for much larger values of
$\mI$ than would be possible for the $\muB$ case, and it is precisely for these large values that bias can become significant. Lastly, the QCD
phase diagram in the $T$-$\mI$ plane is known from several studies to be interesting in its own 
right~\cite{Son:2000xc,Endrodi:2014lja,Brandt:2017oyy,Adhikari:2020kdn,Adhikari:2020ufo}, and we hope to be able to apply this formalism to its study in the future.

\section{Formalism}
\label{sec:formalism}
We consider Lattice QCD with 2+1 flavors of rooted staggered quarks. The partition function at non-zero isospin chemical potential $\mI$ is given by 
\begin{equation}
    Z(T,\mI) = \int \mathcal{D}U e^{-S_G(T)} \,\det M(T,\mI),
\label{eq:partition_function}
\end{equation}
where $\det M(T,\mI)$ is shorthand for
\begin{equation}
\det M(T,\mI) = \prod_{f=u,d,s}\big[\det M_f(m_f,T,\mu_f)\big]^{1/4},
\label{eq:determinant}
\end{equation}
with $m_u=m_d$, $\mu_u=-\mu_d=\mI$ and $\mu_s=0$. The excess pressure $\Delta P(T,\mI) \equiv P(T,\mI) - P(T,0)$ is given by
\begin{equation}
    \frac{\Delta P(T,\mI)}{T^4} = \frac{1}{VT^3} \, \ln \left[\frac{Z(T,\mI)}{Z(T,0)}\right],
\label{eq:excess_pressure}
\end{equation}
where $V$ is the spatial volume and $T$ is the temperature. By employing the same arguments as in Ref.~\cite{Mondal:2021jxk}, we can write
\begin{equation}
    \frac{Z(T,\mI)}{Z(T,0)} = \Bigg \langle
    \exp\left[\sum_{n=1}^\infty \frac{D_{2n}^I(T)}{(2n)!}\mT^{2n}\right] \Bigg \rangle,
    \label{eq:allorders}
\end{equation}
where the expectation value $\langle\cdot\rangle$ is taken over a gauge field ensemble generated at $\mu_u=\mu_d=\mu_s=0$, and
\begin{equation}
    D^I_{n}(T) = \frac{\partial^n \left[\ln\det M(\mI)\right]}{\partial(\mu_I/T)^n}\Bigg\rvert_{\mI=0}.
\label{eq:Dn}
\end{equation}
The presence of only even powers is because the odd $\mI$ derivatives vanish identically. Since even derivatives of the quark determinant are purely real, we see that $\det M(\mI)$ is purely real and hence there is no Sign Problem. Note that this is true even when $\mI$ is purely imaginary.

The $D_n^I$ can be expressed as traces of various operators~\cite{Allton:2005gk,Gavai:2004sd}. In Lattice calculations, the first $N$ derivatives $D^I_1,\dots,D_N^I$ are calculated stochastically using $\Nrv$ random vectors, where $\Nrv$ is typically of order 10$^2$ - 10$^3$. Then $\dP(T,\mI)/T^4$ is
approximately equal to
\begin{equation}
    \frac{\Delta P_N^R(T,\mI)}{T^4} = \frac{\Nt^3}{\Ns^3} \ln 
    \Bigg\langle \exp\left[\sum_{n=1}^{N/2} \frac{\D_{2n}^I(T)}{(2n)!}\mT^{2n}\right]\Bigg\rangle.
\label{eq:resummation}
\end{equation}
Here $\Ns$ and $\Nt$ are the number of lattice sites in the space and time directions respectively, while $\D^I_{2n}$ is the average of the $\Nrv$ stochastic
estimates of $D^I_{2n}$. Eq.~\eqref{eq:resummation} is the $N^\text{th}$ order exponential resummation formula for $\dP(T,\mI)/T^4$. In the limit
$\Nrv\to\infty$, it accurately resums the contribution of the first $N$ derivatives $D_1^I,\dots,D_N^I$ to all orders in $\mI$~\cite{Mondal:2021jxk}. For
$\Nrv<\infty$ however, the formula contains bias. This is easily seen if one writes the exponential as an infinite series. The series expansion leads to terms
such as $(\D^I_{2m})^p (\D^I_{2n})^q\cdots$, and we have already seen that such products are biased due to multiplication of estimates coming from
the same random vector.

The well-known cumulant expansion formula from statistics states that
\begin{equation}
    \ln \big\langle e^{tX} \big\rangle = \sum_{k=1}^\infty \frac{t^k}{k!}\,\C_k(X).
\label{eq:cumulant_expansion}
\end{equation}
The coefficients $\C_k(X)$ are known as the cumulants of $X$~\cite{Endres:2011jm,Ejiri:2007ga,Ejiri:2009hq}. The first four cumulants are given
by
\begin{align}
    \C_1(X) &= \langle X \rangle, \notag \\
    \C_2(X) &= \langle X^2 \rangle - \langle X \rangle^2, \notag \\
    \C_3(X) &= \langle X^3 \rangle - 3 \langle X \rangle \langle X^2 \rangle + 2 \langle X \rangle^3, \notag \\
    \C_4(X) &= \langle X^4 \rangle - 4 \langle X \rangle \langle X^3 \rangle - 3 \langle X^2 \rangle^2 \notag \\
            &+ 12 \langle X^2 \rangle \langle X \rangle^2 - 6 \langle X \rangle^4.
\label{eq:first_four_cumulants}
\end{align}
In our case $t=1$, which we assume lies within the radius of convergence of the cumulant expansion, and $X\equiv X_N(T,\mI)$, where
\begin{equation}
    X_N(T,\mI) = \sum_{n=1}^{N/2} \frac{D_{2n}^I(T)}{(2n)!}\mT^{2n}.
\label{eq:exp_arg}
\end{equation}
Truncating Eq.~\eqref{eq:cumulant_expansion} at $k=M \ge N/2$ \footnote{We have $M\ge N/2$ instead of $M \ge N$ since the first non-vanishing isospin derivative is $D_2^I$ rather than $D_1^I$.} gives us yet another way to estimate $\dP/T^4$, namely
\begin{equation}
    \frac{\dP^C_{N,M}(T,\mI)}{T^4} = \frac{\Nt^3}{\Ns^3}\; \sum_{k=1}^{M \ge N/2} \frac{1}{k!}\, \mathcal{C}_k\big(X_N(T,\mI)\big).
\label{eq:cumulant_pressure}
\end{equation}
Eq.~\eqref{eq:cumulant_pressure} may be compared to the familiar Taylor series expansion of $\dP/T^4$, which in our case is given by
\begin{equation}
    \frac{\Delta P^E_N(T,\mI)}{T^4} = \sum_{n=1}^{N/2} \frac{\chi_{2n}^I(T)}{(2n)!}\mT^{2n}.
\label{eq:QNS_pressure}
\end{equation}
The restriction $M\ge N/2$ in Eq.~\eqref{eq:cumulant_pressure} ensures that the cumulant and Taylor expansions of the pressure agree term-by-term up to 
$\mathcal{O}(\mI^N)$. However, the cumulant expansion also contains additional terms proportional to $\mI^{N+2},\dots,\mI^{MN}$. These extra terms are the same terms that
appear in the calculation of the higher order Taylor coefficients $\chi_{N+2}^I,\dots,\chi_{MN}^I$. The cumulant expansion thus manages to capture some of the higher
order contributions to $\dP/T^4$, although it is not an all orders resummation like Eq.~\eqref{eq:resummation}.
Unlike Eq.~\eqref{eq:resummation} however, only finite products of traces appear in Eq.~\eqref{eq:cumulant_pressure}. As we show in the Appendix, there exist 
formulas (Eqs.~\eqref{eq:girard-newton}~and~\eqref{eq:unbiased_multi})
for evaluating these products efficiently in an unbiased manner. Thus, the cumulant expansion is free of the bias that can affect exponential resummation.

Finally, we will also present results for the net isospin density $\N(T,\mI)$ which is given by
\begin{equation}
    \frac{\N(T,\mI)}{T^3} = \frac{\partial}{\partial \left(\mu_I/T\right)}\left[\frac{\dP(T,\mu_I)}{T^4}\right].
\end{equation}
The Taylor series expression $\N^E_N(T,\mI)$ for the same is straightforward. The resummed and cumulant expansion expressions $\N^R_N(T,\mI)$ and 
$\N^C_{N,M}(T,\mI)$ can be obtained by differentiating Eqs.~\eqref{eq:resummation} and \eqref{eq:cumulant_pressure} respectively. We do not write down the explicit expressions here. Note however that the resummed formula, unlike the cumulant expansion expression, involves a ratio of expectation values.

\section{Results}
\label{sec:results}
\begin{figure}[!tb]
\includegraphics[width=0.45\textwidth]{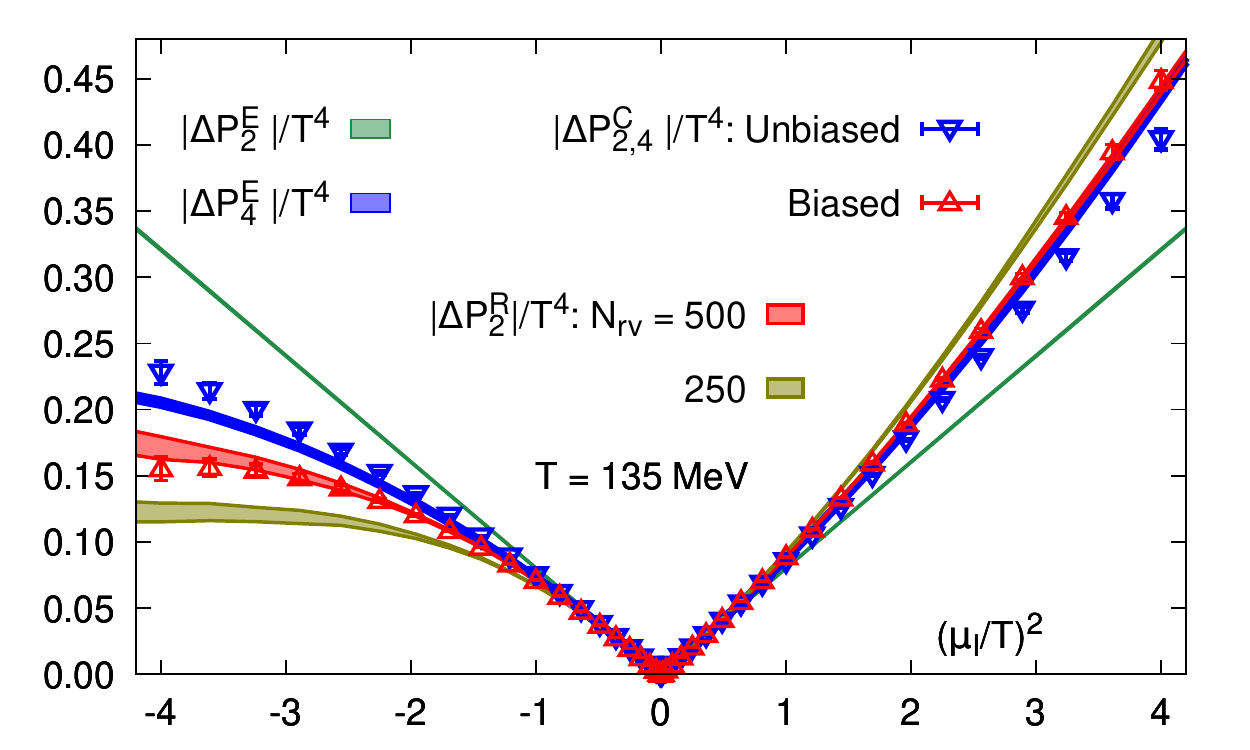}
\includegraphics[width=0.45\textwidth]{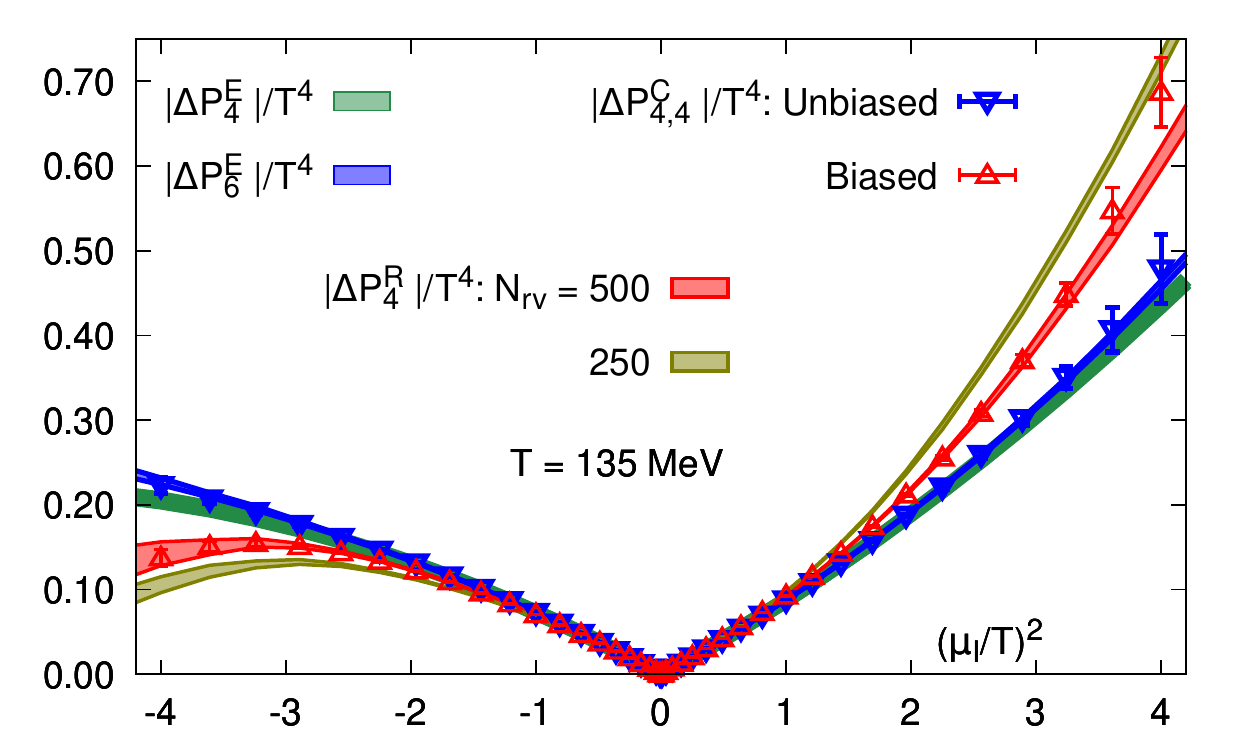}
\caption{(Color online) Comparison of the results for the excess pressure $\dP(T,\mI)$ obtained using Taylor series expansion, exponential resummation and
cumulant expansion. The resummed results were obtained for $\Nrv=500$ and $\Nrv=250$ (red and yellow bands). The cumulant expansion was calculated
using both biased  and unbiased estimates (upright and inverted triangles).\\
(Top)~Comparison between $\second$ and $\fourth$ order Taylor expansions, $\second$ order resummation and $(N,M)=(2,4)$ cumulant expansion. 
(Bottom)~Comparison between $\fourth$ and $\sixth$ order Taylor expansions, $\fourth$ order resummation and $(N,M)~=~(4,4)$ cumulant expansion.}
\label{fig:PT4_second_fourth}
\end{figure}

To verify our formalism, we made use of the data generated by the HotQCD collaboration for their calculations of the finite-density EoS, finite-density chiral
crossover temperature and conserved charge cumulants using Taylor series expansions~\cite{Bazavov:2017dus,Bazavov:2018mes,Bollweg:2022rps}. The data
consists of 2+1-flavor gauge configurations with $\Nt=8$, 12 or 16 and $\Ns=4\Nt$ in the temperature range 125~MeV~$\lesssim T\lesssim$~178~MeV. The 
configurations were generated using a Symanzik-improved gauge action and the Highly Improved Staggered Quark (HISQ)
action~\cite{Follana:2006rc,Bazavov:2011nk,Bazavov:2014pvz} for fermions. The lattice spacing was determined using both the Sommer parameter $r_1$ as well as the
decay constant $f_K$. The temperature values quoted in this paper were obtained using the $f_K$ scale. For each lattice spacing, the light and strange quark bare 
masses were tuned so that the pseudo-Goldstone meson masses reproduced the physical pion and kaon masses. A description of the gauge ensembles and scale setting 
can be found in Ref.~\cite{Bollweg:2021vqf}.

The results presented here were obtained with around 20,000 $\Nt=8$ configurations for $T=135$~MeV. On each gauge configuration, the first eight 
derivatives $D_1^f,\dots,D_8^f$ for each quark flavor were estimated stochastically using around 2000 Gaussian random volume sources for $D^f_1$ and around 500
sources for the rest. We used the exponential-$\mu$ formalism~\cite{Hasenfratz:1983ba} to calculate the first four derivatives, while the
linear-$\mu$ formalism~\cite{Gavai:2011uk,Gavai:2014lia} was used in calculating all higher derivatives.

We calculated the excess pressure $\dP(T,\mI)$ and net isospin density $\N(T,\mI)$ using (i)~Taylor series expansion (Eq.~\eqref{eq:QNS_pressure}) 
for $N=2,4$ and 6; (ii)~exponential resummation (Eq.~\eqref{eq:resummation}) for $N=2$ and 4; and (iii)~a fourth order cumulant expansion 
(Eq.~\eqref{eq:cumulant_pressure} with $M=4$) for $N=2$ and 4. We calculated these observables for both real as well as imaginary $\mI$, in the range $0 \leqslant \lvert
\mI/T \rvert \leqslant 2$.

\begin{figure}[!tb]
\includegraphics[width=0.45\textwidth]{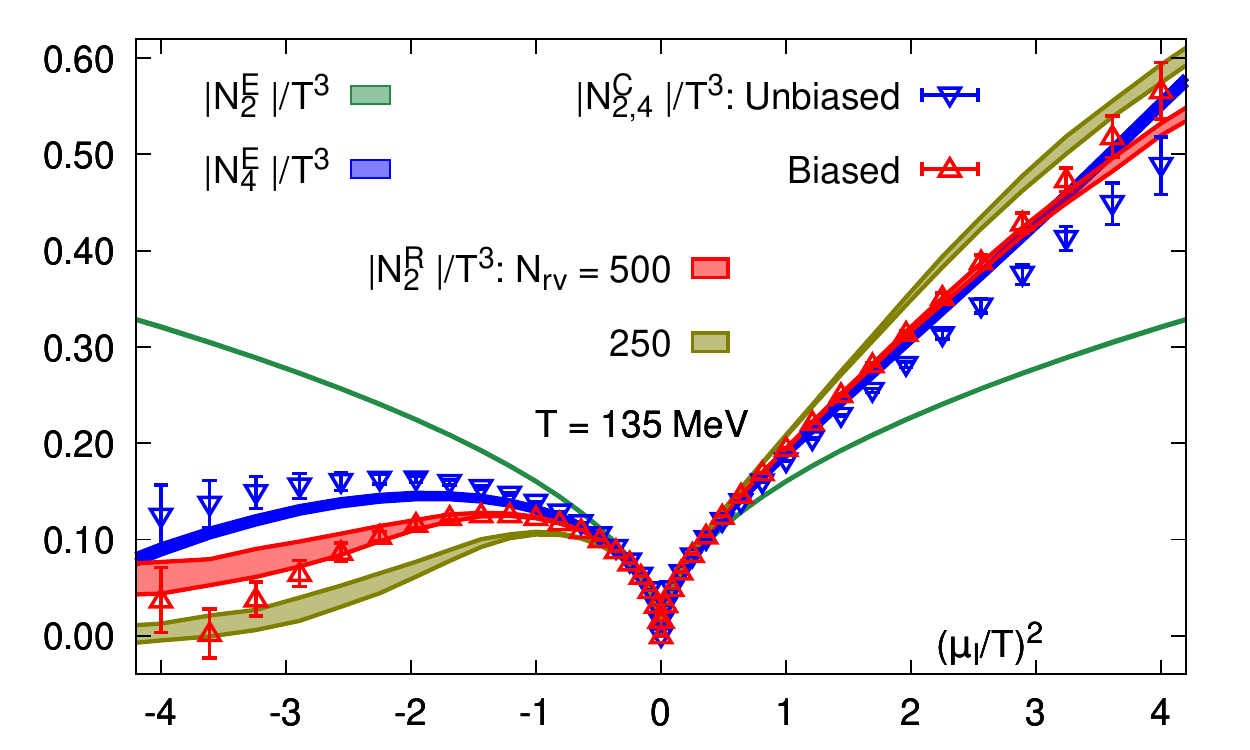}
\includegraphics[width=0.45\textwidth]{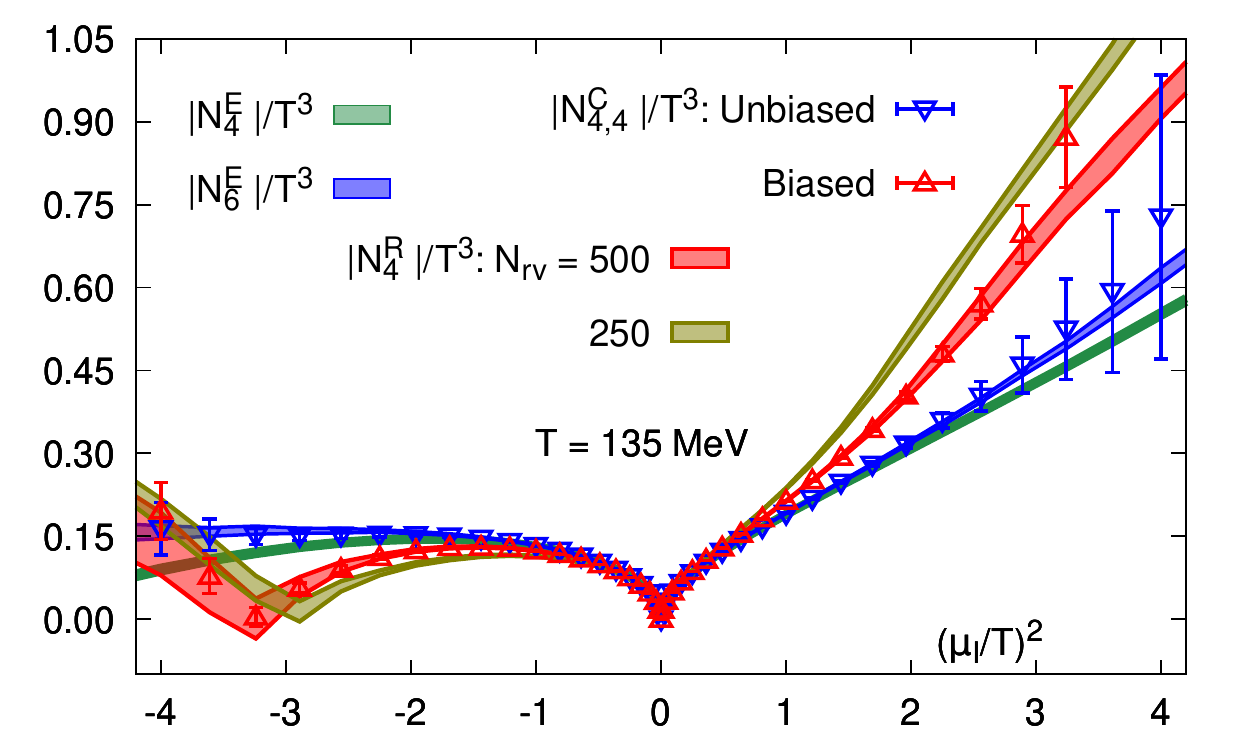}
\caption{(Color online) Comparison of the results for the isospin density $\N(T,\mI)$ obtained using Taylor series expansion, exponential resummation and
cumulant expansion.  All colors and symbols are the same as in Fig.~\ref{fig:PT4_second_fourth}.\\
(Top)~Comparison between $\second$ and $\fourth$ order Taylor expansions, $\second$ order resummation and $(N,M)=(2,4)$ cumulant expansion. 
(Bottom)~Comparison between $\fourth$ and $\sixth$ order Taylor expansions, $\fourth$ order resummation and $(N,M)~=~(4,4)$ cumulant expansion.}
\label{fig:NT3_second_fourth}
\end{figure}

We present our results for $\dP(T,\mI)$ and $\N(T,\mI)$ in Figs.~\ref{fig:PT4_second_fourth} and \ref{fig:NT3_second_fourth} respectively.
The upper plots in each figure compare second order resummation and an $(N,M)=(2,4)$ cumulant expansion results with second and fourth order Taylor expansions, while
the lower plots compare fourth order resummation and $(N,M)=(4,4)$ cumulant expansion results with fourth and sixth order Taylor expansions respectively.

Focussing first on the upper plots, we see that the second and fourth order Taylor expansion results (green and blue bands)
start to differ significantly around $\lvert\mu_I/T\rvert=1$. For real $\mI$, this difference is seemingly captured by the resummed results (red band) which
agree with the fourth order Taylor results for both observables over nearly the entire range of $\mI/T$. For imaginary $\mI$ however, the resummed results
lie below the fourth order Taylor results. By contrast, the cumulant expansion results (blue inverted triangles) are in good agreement with the fourth order Taylor 
results for imaginary $\mI$, while they are only slightly less than the fourth order Taylor results for real $\mI$.

One explanation for the difference between the resummed and cumulant results is the higher order contributions that are present in the former but not in the
latter. Another possibility is the bias that is present in the resummed but not in the cumulant results. To distinguish between the two possibilities, we
recalculated the cumulant results using the biased formulas for the trace products rather than the unbiased ones (Eq.~\eqref{eq:biased} rather than
Eq.~\eqref{eq:unbiased_formal}). We found that the biased results (upright red triangles) agree well with the resummed results, thus suggesting that bias, rather
than the contribution from higher orders, is responsible for the difference between the resummed and the cumulant expansion results.

To further confirm that this is the case, we also recalculated the resummed results using only 250 random vectors instead of 500. Since bias decreases as $\Nrv$ is
increased, conversely we should expect the bias to increase when we use fewer random vectors. From Figs.~\ref{fig:PT4_second_fourth} and \ref{fig:NT3_second_fourth}, we
see that the $\Nrv=250$ results (yellow band) lie further from the Taylor and unbiased cumulant results than the $\Nrv=500$ results for $\lvert\mI/T\rvert\gtrsim1$. 
Thus we see that the resummed results are indeed affected by bias for large values of the chemical potential.

The presence of bias must especially be accounted for when comparing higher order results (lower plots in Figs.~\ref{fig:PT4_second_fourth} and
\ref{fig:NT3_second_fourth}). We see that the sixth order Taylor correction (blue band) to the fourth order results (green band) is small over the entire range of $\mI$
considered here. By contrast the resummed results, although not containing the contributions of the operator $D^I_6$, nonetheless suggest that the higher-order 
contributions of $D_2^I$ and $D_4^I$ are large for both real as well as imaginary $\mI$. However, both the biased cumulant expansion results as well as the $\Nrv=250$
resummed results once again suggest that at least some of this difference is due to bias. On the other hand, the unbiased cumulant results agree well with the Taylor
series results for both real and imaginary $\mI$. We note that while the cumulant expansion too does not include the contribution of the operator $D_6^I$, it does contain
higher order corrections that appear at $\O(\mI^6,\dots,\mI^{16})$. In fact, the $(4,4)$ cumulant expansion is exactly equal to a fourth order Taylor expansion, plus the
contributions of the operators $D_2^I$ and $D_4^I$ to the Taylor coefficients $\chi^I_6,\dots,\chi^I_{16}$. The agreement between the unbiased cumulant and Taylor series
results thus suggests that the contribution of $D_2^I$ and $D_4^I$ at higher orders is in fact small. All this goes to show that bias needs to be properly accounted for
before one can identify the genuine higher-order corrections.

\section{Conclusions}
\label{sec:conclusions}
In this paper, we presented a new way of resumming the QCD Taylor series EoS based on the well-known cumulant expansion from statistics. Our approach is a
finite order truncation of the all orders resummation of the first $N$ charge density correlation functions that was presented in Ref.~\cite{Mondal:2021jxk}. 
The resummation presented there is susceptible to bias when the correlation functions are calculated stochastically. By contrast, the cumulant expansion
contains only finite products of traces that can be evaluated efficiently in an unbiased manner. Moreover, while not an all orders resummation, the $M$th order
cumulant expansion still captures the contributions of the lower order derivatives $D_1,\dots,D_N$ to the higher order Taylor coefficients up to a maximum 
order $\chi_{MN}$.

Although our formalism is also applicable to $\muB\ne0$, in this paper we presented results for finite isospin chemical potential $\mI$ instead. Our reason for this was that the absence of a Sign Problem in the latter case meant that all quantities were real and this simplified the presentation. The $\mI\ne0$ case is
also of interest in its own right. We presented results for the excess pressure and net isospin density using Taylor series, resummation and cumulant expansion.
We found evidence for bias in the resummed results at large $(\mI/T)^2$. We showed this by (a) calculating the cumulant expansion using biased rather than
unbiased products, and (b) recalculating the resummed results using fewer random vectors. The cumulant expansion is a truncation of the resummed formula and when
the terms of the expansion were calculated using biased products of traces, they were in good agreement with the resummed result, while they were closer to the
Taylor series results when they were calculated using unbiased products.

There have been several proposals recently to extend the QCD EoS to larger values of the chemical potential~\cite{Borsanyi:2021sxv,Bollweg:2022rps,Dimopoulos:2021vrk,Pasztor:2020dur,Giordano:2020roi}. Exponential resummation is one such approach, which
can be connected to reweighting-based approaches. By generalizing it to resummation in $T$ and $\muB$, it can be also connected with the alternative expansion
scheme proposed in  Ref.~\cite{Borsanyi:2021sxv}. The cumulant expansion approach that we have outlined here provides yet another way to extend the QCD EoS. 
The precise relation between this approach and the other proposals remains to be studied.

All data presented in the figures of this paper can be found in Ref.~\cite{Schmidt:2022bf}.

\acknowledgments
We thank all the members of the HotQCD collaboration for their inputs and for valuable discussions, as well as for allowing us to use their data from the Taylor expansion calculations. CS was supported by the European Union’s Horizon 2020 research and innovation program under the Marie Sklodowska-Curie
Grant Agreement No. H2020-MSCAITN-2018-813942 (EuroPLEx) and The Deutsche Forschungsgemeinschaft (DFG, German Research Foundation), Project Number 315477589-TRR 211. This work made use of the computing and data storage facilities, as well as the GPU-cluster, at Bielefeld University, Germany. We are very grateful to the Bielefeld HPC.NRW team for their help and support.

\appendix*
\section{On the Calculation of Unbiased Estimators of Powers of the Trace of an Operator}
\label{sec:unbiased}
Consider a set of $\Nrv$ independent and identically distributed random estimates $\O_1,\O_2,\dots,\O_\Nrv$ of the trace of an operator $\O$
\footnote{Estimates $\O_i$ of the trace of an operator $\O$ are usually obtained by drawing a set of random volume source vectors $\eta_i$, with components fulfilling the normalization condition $\lim_{\Nrv\to\infty} N_\text{rv}^{-1} \sum_{i=1}^\Nrv\eta_{i,x}^\dagger\eta^{\phantom{\dagger}}_{i,y}=\delta_{x,y}$, where $x$ and $y$ are lattice sites. The scalars $\O_i$ are then defined as $\O_i=\eta_i^\dagger\O\eta_i^{\phantom{\dagger}} \equiv \sum_{x,y} \eta_{i,x}^\dagger\,\O^{\phantom{\dagger}}_{i,x,y}\,\eta^{\phantom{\dagger}}_{i,y}$.}. We have
\begin{equation}
    \tr\,\O = \lim_{\Nrv \to \infty} \frac{1}{\Nrv}\sum_{i=1}^\Nrv \, \O_i.
\end{equation}
In practice, the limit $\Nrv\to\infty$ is usually dropped and the trace is estimated by the arithmetic mean over a finite number of estimates $\O_i$, which bears potential danger for the estimation of powers of the trace $(\tr\,\O)^n$, with $n\ge 1$. The na\"ive estimate 
\begin{equation}
    \text{BE}\left[(\tr\,\O)^n\right] \equiv \left[\frac{1}{\Nrv}\sum_{i=1}^\Nrv \O_i\right]^n,
\label{eq:biased}
\end{equation}
is a biased estimate since it receives contributions from products of the same estimates $\O_i^k, k\le n$, which lead to a finite bias in the estimator
\eqref{eq:biased}. The bias can be quite significant in the regime of large $n$, even when we take $\Nrv\sim\O(1000)$. An unbiased estimator for $(\tr\,\O)^n$ can be
obtained by rejecting all diagonal contributions and considering only mutually distinct estimates in each product. We define 
\begin{align}
    \text{UE}\left[\left(\tr\,\O\right)^n\right] & \equiv \frac{n!}{\prod_{k=0}^{n-1}(\Nrv-k)}\sum_{i_1=1}^{\Nrv}\quad\cdots\sum_{i_n=i_{n-1}+1}^{\Nrv} \O_{i_1}\dots\O_{i_n}\notag\\
    &=\frac{n!}{\prod_{k=0}^{n-1}(\Nrv-k)}\,e_n(\O_1,\O_2,\dots,\O_n)\,,
\label{eq:unbiased_formal}
\end{align}
where $e_n$ denotes the \textit{elementary symmetric polynomial} of degree $n \le \Nrv$, in $\Nrv$ variables. For later use, we define $e_0(\O_i,\dots\O_\Nrv)\equiv 1$.

Although Eq.~\eqref{eq:unbiased_formal} provides us with a valid unbiased estimator, its evaluation requires $\O(N_\text{rv}^n)$ elementary operations due to the nested sums. Fortunately, it is possible to rearrange the above sum so that it can be evaluated in only $\O(\Nrv)$ operations by utilizing the \textit{Girard-Newton identities}, which relate the elementary symmetric polynomials to the \textit{power sums} $p_k(\O_1,\O_2,\dots,\O_\Nrv)=\sum_{i=1}^{\Nrv}\O_i^k$. Suppressing the arguments of the polynomials, the Girard-Newton identities read 
\begin{align}
ne_n=\sum_{k=1}^{n}(-1)^{k-1}e_{n-k}p_k\;,
\label{eq:girard-newton}
\end{align}
which can be readily used to obtain a recursive schema for the evaluation of the unbiased estimators in Eq.~\eqref{eq:unbiased_formal}. We can also use the tower of identities~\eqref{eq:girard-newton} to completely erase the elementary symmetric polynomials from the right-hand side and express the unbiased estimator by power sums only,
\begin{align}
&\text{UE}\left[\left(\tr\,\O\right)^n\right] = \frac{(-1)^n\,B_n\big(-p_1,-1!\,p_2,\dots,-(n-1)!\,p_n\big)}{\prod_{k=0}^{n-1}(\Nrv-k)}, \notag \\
&\frac{1}{n!}\,B_n\left(p_1,p_2,\dots,p_n\right)\,=\sum_{\substack{k_1+\cdots+nk_n=n \\ k_1,\dots,k_n\ge 0}}\,\,
\prod_{i=1}^n\frac{1}{k_i!}\left[\frac{p_i}{i!}\right]^{k_i}.
\label{eq:unbiased_powersum_expression}
\end{align}
In this case the sum is taken over all partitions of $n$, which makes the connection to the \textit{complete exponential Bell polynomials} $B_n$ and is made explicit with the second equality.  Below we give the final formulae in the form of Eq.~\eqref{eq:unbiased_powersum_expression}, which are required up to $n=4$,
\begin{align}
    \text{UE}\left[\left(\trO\right)^2\right] &= \frac{p_1^2-p_2}{\Nrv(\Nrv-1)}\,,\notag\\
    \text{UE}\left[\left(\trO\right)^3\right] &= \frac{p_1^3-3p_1p_2+2p_3}{\Nrv(\Nrv-1)(\Nrv-2)}\,,\notag\\
    \text{UE}\left[\left(\trO\right)^4\right] &= \frac{p_1^4-6p_1^2p_2+3p_2^2+8p_1p_3-6p_4}{\Nrv(\Nrv-1)(\Nrv-2)(\Nrv-3)}.
\label{eq:unbiased_actual}
\end{align} 
Although there are several (power) sums to be evaluated, these are not nested sums as in Eq.~\eqref{eq:unbiased_formal}.
Therefore, the unbiased product can be calculated in only $\O(\Nrv)$ time.

\subsection*{Unbiased Estimators for Combinations of Traces of Multiple Operators}
Quite often we encounter a situation where we need to estimate expressions involving combinations of traces of several operators $\O^{(1)}, \O^{(2)}, \dots, \O^{(m)}$, which are estimated on the same set of random vectors and are thus correlated. 
The evaluation of different operators on the same set of random vectors might seem to be avoidable at the first glance but  could -- as in the case of the derivative operators of the pressure discussed above -- gain computational advantages, e.g., due to a recursive definition of the 
operators~\cite{Allton:2005gk,Steinbrecher:2018jbv}.
A biased estimator of a general expression of this kind is given by 
\begin{align}
\text{BE}&\left[\left( \tr\,\O^{(1)}\right)^{\gamma_1}\; \left(\tr\,\O^{(2)}\right)^{\gamma_2}  \cdots \left(\tr\,\O^{(m)}\right)^{\gamma_m}\right] \notag \\
&=\frac{1}{N_\text{rv}^{\sum_i \gamma_i}}\left\{\sum_{k=1}^\Nrv\O^{(1)}_k\right\}^{\gamma_1}\cdots\left\{\sum_{k=1}^\Nrv\O^{(m)}_k\right\}^{\gamma_m}.
\label{eq:muli_operator_biased}
\end{align}
In order to construct an unbiased estimator for this more general case, we extend the framework of elementary symmetric polynomials and power sums presented above as follows: We introduce the multi-index notation $\alpha, \beta, \gamma\in \mathbb{N}^m$ with non-negative integer coefficients and define the metric $|\alpha|\equiv \sum_i\alpha_i$. We further define the two types of symmetric polynomials as 
\begin{align}
p_\alpha&=\sum_{i=1}^\Nrv \left\{\O^{(1)}_i\right\}^{\alpha_1} \left\{\O^{(2)}_i\right\}^{\alpha_2} \dots \left\{\O^{(m)}_i\right\}^{\alpha_m}\notag\\
e_\beta&=\sum_{i_1<i_2<\cdots<i_{|\beta|}} \underbrace{\O^{(1)}_{i_1}\cdots \O^{(1)}_{i_{\beta_1}}}_{\beta_1\text{-times}}\;
\underbrace{\O^{(2)}_{i_{\beta_1+1}}\cdots \O^{(2)}_{i_{\beta_1+\beta_2}}}_{\beta_2\text{-times}} \cdots \notag \\
&\phantom{=sum_{i_1<i_2<\cdots<i_{|\beta|}}}\underbrace{\O^{(m)}_{i_{\beta_1+\cdots + \beta_{m-1}+1}} \cdots \O^{(m)}_{i_{|\beta|}}}_{\beta_m\text{-times}}\,.
\end{align}
The analogues of the Girard-Newton identities \eqref{eq:girard-newton} are than given as 
\begin{equation}
|\gamma|\cdot e_\gamma=\sum_{\alpha+\beta=\gamma, |\alpha|>0} (-1)^{|\alpha|-1}\binom{|\alpha|}{\alpha}p_\alpha e_\beta\,,
\label{eq:girard-newton_multi}
\end{equation}
where the addition of multi-indices $\alpha+\beta$ is understood by components, the summation is over the product of all partitions of the components of $\gamma$ and $\binom{|\alpha|}{\alpha}$  denote the multinomial coefficients. We can now define an unbiased version of \eqref{eq:muli_operator_biased} based on the symmetric polynomial $e_\gamma$ as 
\begin{equation}
\text{UE}\left[ \left( \tr\,\O^{(1)} \right)^{\gamma_1} \cdots \left(\tr\,\O^{(m)}\right)^{\gamma_m} \right] = \frac{|\gamma|!\cdot e_\gamma}{\Nrv\cdots(\Nrv-|\gamma|+1)}.
\label{eq:unbiased_multi}
\end{equation}
For the practical calculation of $e_\gamma$ we use the recursive definition \eqref{eq:girard-newton_multi}. To reduce the computational efforts we manually cache previously computed values of $p_\alpha$ and $e_\beta$. We stress again that the main computational effort goes into the evaluation of the power sums. Each power sum has complexity $\O(\Nrv)$. The number of distinct power sums given by $\prod_i(\gamma_i+1) -1$. Even though the power sum number increases also drastically with the order $|\gamma|$, the values we encounter in this calculation are still of the order $\O(100)$ and thus an order of magnitude lower than the costs of a single power sum evaluation. For convenience we give in Tab.~\ref{tab:partitions} some examples for cases with $|\gamma|=8$, which we identified as the cases with the maximum number of power sums, for a given number of operators $m$. Note that for three-quark flavors ($u,d,s$) and $|\gamma|\le 8$, the maximum number of distinct operators we encounter in and unbiased estimator is $m=5$.
\begin{table}[!htb]
\begin{center}
\begin{tabular}{c|c|c|c|c|c}
$m$ & 1 & 2 & 3 & 4 & 5   \\ \hline 
$\gamma_{\text{max}}$ & $(8)$ & $(4,4)$ & $(3,3,2)$ & $(2,2,2,2)$ & $(2,2,2,1,1)$  \\ \hline
\# distinct power sums & 8 & 24 & 47 & 80 & 107 
\end{tabular}
\end{center}
\caption{The maximum number of distinct power sums that need to be evaluated for an unbiased estimator of order $|\gamma|=8$, as a function of the number of distinct operators $m$ (dimension of $\gamma$). \label{tab:partitions}}
\end{table}

\bibliographystyle{apsrev4-2.bst}
%

\end{document}